\documentclass{PoS}

\usepackage{amssymb, amsmath, cite} 
\usepackage{float}

\title{{\footnotesize 
DESY 14--130, DO-TH 14/14, MITP/14-048, SFB/CPP-14-53, LPN14--091}\\
Recent progress on the calculation of three-loop heavy flavor Wilson coefficients in deep-inelastic 
scattering}

\ShortTitle{Recent progress on the calculation of three-loop heavy flavor Wilson coefficients in DIS}

\author{J.~Ablinger$^a$, A.~Behring$^b$, J.~Bl\"umlein$^b$, \speaker{A.~De~Freitas}$^b$\thanks{
We would like to thank M.~Steinhauser for the possibility to use the code {\tt MATAD}. This work was supported
in part by DFG Sonderforschungsbereich Transregio 9, Computergest\"utzte Theoretische Teilchenphysik, Studienstiftung
des Deutschen Volkes, the Austrian Science Fund (FWF) grants P20347-N18 and SFB F50 (F5009-N15), the European
Commission through contract PITN-GA-2010-264564 ({LHCPhenoNet}) and PITN-GA-2012-316704 ({HIGGSTOOLS}), by the
Research Center ``Elementary Forces and Mathematical Foundations (EMG)'' of J. Gutenberg University Mainz and
DFG, and by FP7 ERC Starting Grant  257638 PAGAP.},
        A.~Hasselhuhn$^a$, A.~von~Manteuffel$^c$, C.~Raab$^b$, M.~Round$^b$, C.~Schneider$^a$, and 
        F.~Wi\ss{}brock$^{a,b,d}$ 
\\
        \llap{$^a$}Research Institute for Symbolic Computation (RISC), Johannes Kepler University, \\ 
                   Altenbergerstra\ss{}e 69, A--4040, Linz, Austria. \\
        \llap{$^b$}Deutsches Elektronen--Synchrotron, DESY, \\
                   Platanenallee 6, D-15738 Zeuthen, Germany. \\
        \llap{$^c$}PRISMA Cluster of Excellence, Institute of Physics, J. Gutenberg University, \\
                   D-55099 Mainz, Germany.\\
        \llap{$^d$}~Institut des Hautes {\'E}tudes Scientifiques, IHES, Route de Chartres 35, \\ F--91440 Bures-sur-Yvette, 
        France\\
E-mails: 
\email{jablinge@risc.uni-linz.ac.at},
\email{Arnd.Behring@desy.de},
\email{Johannes.Bluemlein@desy.de},
\email{abilio.de.freitas@desy.de},
\email{alexander.hasselhuhn@desy.de},
\email{manteuffel@uni-mainz.de},
\email{clemens.raab@desy.de},
\email{mark.round@desy.de},
\email{cschneid@risc.uni-linz.ac.at},
\email{fabian.wissbrock@desy.de} 
}

\abstract{We report on our latest results in the calculation of the three-loop heavy flavor contributions to the Wilson
coefficients in deep-inelastic scattering in the asymptotic region $Q^2 \gg m^2$. We discuss the different methods 
used to compute the required operator matrix elements and the corresponding Feynman integrals. These methods very 
recently allowed us to obtain a series of new operator matrix elements and Wilson coefficients like the flavor non-singlet 
and pure singlet Wilson coefficients.}

\FullConference{ Loops and Legs in Quantum Field Theory - LL 2014,\\
		 27 April - 2 May 2014 \\
		 Weimar, Germany }

\begin{document}

\section{Introduction}

\vspace*{1mm}
\noindent
In the large $Q^2$ limit, the heavy flavor Wilson coefficients in deep-inelastic scattering (DIS) are known to factorize into light
flavor Wilson coefficients and massive operator matrix elements \cite{BMSN96,BBK09NPB}. These heavy flavor 
coefficients can then be convoluted with parton distribution functions (PDFs) to obtain the heavy flavor corrections 
to deep-inelastic scattering structure functions, which amount to sizeable contributions, in particular 
in the region of small values of the Bjorken variable $x$. At NNLO, the light flavor Wilson coefficients are 
known \cite{MVV2005}. The missing ingredients required to obtain the heavy flavor Wilson coefficients 
are therefore the 3-loop massive operator matrix elements (OMEs). 

Here we present our latest results in the ongoing effort to calculate these quantities. Particularly, six out of eight OMEs 
are by now available. The simplest operator matrix elements, $A_{qq,Q}^{(3), \rm PS}$ and $A_{qg,Q}^{(3)}$, were obtained 
in \cite{ABKSW11NPB}. 
More recently, $A_{gq}^{(3)}$ \cite{Ablinger:2014lka}, $A_{qq}^{(3), \rm NS, TR}$ \cite{Ablinger:2014vwa} and 
$A_{Qq}^{(3), \rm PS}$ \cite{PS} were calculated using a variety of techniques to be described in the following 
sections. Using these operator matrix elements, we have been able to obtain the Wilson 
coefficients $L_{q,(2,L)}^{\rm PS}$, $L_{g,(2,L)}^{\rm S}$ \cite{Behring:2014eya}, 
$L_{q,(2,L)}^{\rm NS}$ \cite{Ablinger:2014vwa} and $H_{q,(2,L)}^{\rm PS}$ \cite{PS}.
The asymptotic heavy flavor corrections to $F_L(x,Q^2)$ have been calculated in Refs.~\cite{Blumlein:2006mh,Behring:2014eya}.
Also, results for the terms proportional to $T_F^2$ in $A_{gg,Q}^{(3)}$ \cite{Ablinger:2014uka}
and all $N_F T_F^2$ terms   \cite{ABKSW11NPB,Blumlein:2012vq,Behring:2013dga}
have been computed. Once the last two operator matrix elements are 
completed, and the corresponding heavy flavor contributions to the DIS structure functions are then obtained,
it will be possible to make more precise determinations of $\alpha_s$ and the mass of the charm quark $m_c$, 
as well as provide better constraints on sea quarks and the gluon and thus improve the results given in~\cite{Alekhin:2012vu,PDF,Alekhin:2013nda}.
The 3-loop OMEs are also needed to obtain the matching relations at NNLO in the variable flavor number scheme (VFNS) 
\cite{Buza:1996wv,BBK09NPB}. Starting with 3-loop order there are also heavy flavor contributions due to graphs containing 
massive fermion lines of different mass, see Refs.\cite{Ablinger:2011pb,Ablinger:2012qj,JB14a} and the talk by F.~Wi\ss{}brock 
\cite{Ablinger:2014fla} presented at this conference. Furthermore, the asymptotic heavy flavor corrections 
for the charged current processes have also been calculated to
next-to-leading order (NLO) \cite{Buza:1997mg,Blumlein:2014fqa}.

The operator matrix elements have been calculated using the standard Feynman rules of QCD together with the 
Feynman rules for operator insertions as described in Refs. \cite{Ablinger:2014vwa,Klein:2009ig}. Feynman 
diagrams were generated based on these rules using {\tt QGRAF} \cite{Nogueira:1991ex}. The output of {\tt QGRAF} 
was then processed using {\tt Form} \cite{FORM}, after which the diagrams end up being expressed as a linear 
combination of a large number of scalar integrals. These scalar integrals are then reduced to a much smaller 
set of master integrals using integration by parts identities, as described in Section~\ref{IBP}. The master 
integrals are then calculated using a variety of techniques. These will be discussed in Section~\ref{MI}, where 
we will make special emphasis on the differential equations method. We will discuss our results in Section~\ref{RES}.
The conclusions are given in Section~\ref{CONC}.
\section{Calculation of the operator matrix elements}

\vspace*{1mm}
\noindent
\subsection{Integration by parts identities}
\label{IBP}

\vspace*{1mm}
\noindent
We calculate the operator matrix elements as functions of the Mellin variable $N$, and perform the reduction to master 
integrals using the {\tt C++} program {\tt Reduze2} \cite{vonManteuffel:2012np}\footnote{The program {\tt Reduze2} uses 
the codes {\tt Fermat}~\cite{FERMAT} and {\tt GiNaC}~\cite{Bauer:2000cp}.}, which implements Laporta's algorithm 
\cite{Laporta:2001dd}. It is not so straightforward to apply this algorithm to the case where we have operator insertions, 
precisely because of the dependence on the arbitrary parameter $N$. Laporta's algorithm is designed to work with integrals 
with definite indexing, and although it may be possible to adapt this algorithm to the case where we have an arbitrary 
index $N$, we found a more elegant solution by rewriting the operator insertions in terms of propagators, which will be 
raised to definite powers. For example, in the case of an insertion on a line, the operator insertion will be proportional to 
$(\Delta.k)^{N-1}$, where $k$ is the momentum going through the line, and $\Delta$ is a light-like vector. We now introduce 
a new variable $x$ and re-express the operator insertion by the following generating function
\begin{equation}
(\Delta.k)^{N-1} \rightarrow \sum_{N=1}^{\infty} x^{N-1} (\Delta.k)^{N-1} = \frac{1}{1 - x \Delta.k}.
\label{trick}
\end{equation}
In the case of 3-, 4- and 5-point operator insertions, we can similarly re-express the operator insertion in terms of 
products of the same type of artificial propagators. These new propagators can be added to the list of propagators 
defining the integrals, and Laporta's algorithm can then be applied without problems. Propagators like the one given 
in Eq.~(\ref{trick}) are known as bilinear propagators, and {\tt Reduze2} has been adapted to be able to deal 
with such objects. Auxiliary propagators are introduced when needed in such a way that all products of internal momenta
with an external/internal momentum or with $\Delta$ can be uniquely expressed as a linear combination of inverse propagators. 
A set of propagators that satisfies this condition is called an {\it integral family}. All integrals involved in a given 
problem will be identified by specifying an integral family and the powers of the propagators, which can be negative if 
the integral has irreducible numerators. We have found that all integrals required for the calculation of all of the eight 
OMEs can be specified using 24 integral families.
\subsection{Calculation of the master integrals}
\label{MI}

\vspace*{1mm}
\noindent
For the calculation of the master integrals we used a combination of one or more of the following methods, depending on 
the complexity of the integral under consideration:
\begin{itemize}
\item Summation methods, implemented in the {\tt Mathematica} package {\tt Sigma} 
\cite{SIG1,SIG2}, based on advanced symbolic summation algorithms in the setting of difference fields
\cite{Karr:81,Schneider:01,Schneider:05a,Schneider:07d,Schneider:08c,Schneider:10a,Schneider:10b,Schneider:10c,
Schneider:13b}, and the packages 
{\tt HarmonicSums} \cite{Ablinger:2010kw,Ablinger:2011te,Ablinger:2013cf,Ablinger:2013hcp},
{\tt EvaluateMulti- Sums}, and {\tt SumProduction} \cite{EMSSP}.
\item Hypergeometric functions \cite{HYPERGEOM,TWOL,Ablinger:2012qm}.
\item Mellin-Barnes representations \cite{Mellin1895,Barnes1908,Smirnov:1999gc,Smirnov2006}.
\item In the case of convergent massive 3-loop Feynman integrals, they can be performed in terms of 
hyperlogarithms, generalizing the method proposed in \cite{Brown:2008um} to massive diagrams with 
operator insertions \cite{Ablinger:2012qm,Ablinger:2014yaa}.
\item Differential (difference) equations \cite{DEQ}.
\end{itemize}

\noindent
The summation methods and other mathematical methods we used in the representation and reduction of nested sums and 
iterated integrals 
we have used were described in detail in the talks by C.~Schneider~\cite{Bluemlein:2014qka}, 
J. Ablinger~\cite{ABLINGER:14} and C. Raab~\cite{RAAB:14,Ablinger:2014bra} given at this conference. Some of the integrals can 
be completely 
solved in terms of hypergeometric functions (including Appell hypergeometric functions) 
with parameters depending on $N$ and the dimension $D = 4+\varepsilon$, or multiple sums of such functions where the summation indices also appear in the parameters
of the hypergeometric function. If the corresponding series representation is convergent, the resulting sums can then be performed 
using {\tt Sigma}. A survey on the function spaces, which have appeared in the present calculations, is given in 
Ref.~\cite{Ablinger:2013jta}. 

In some cases, after Feynman parameterization of the integrals, the Feynman parameters can be integrated in terms of Beta functions by splitting a denominator using \cite{Smirnov:1999gc,Smirnov2006}
\begin{equation}
\frac{1}{(A+B)^{\nu}} = \frac{1}{2 \pi i} \int_{-i \infty}^{+i \infty} d\sigma \, \, \frac{\Gamma(-\sigma) \Gamma(\sigma+\nu)}{\Gamma(\nu)} \frac{A^{\sigma}}{B^{\sigma+\nu}}.
\end{equation}
The remaining contour integral in $\sigma$ can then be done with the help of the {\tt Mathematica} package {\tt MB} 
\cite{Czakon:2005rk}, which finds a contour and a value of $\varepsilon$ such that the
Feynman integral is well defined, and then analytically continues to $\varepsilon \rightarrow 0$. After this, we can take residues and then sum them using {\tt Sigma}.

The method of hyperlogarithms and its generalizations have been described in detail in \cite{Ablinger:2014yaa}, where a few examples were 
presented. This method applies to Feynman integrals which are non-singular in the dimensional parameter $\varepsilon$. It relies on the 
$\alpha$-parameterization of the integrals, integrating each parameter one after the other. A required 
condition for the applicability of this method is that after each integration, the denominators of the integrals 
remain linearly factorizable in the $\alpha$ parameters. Many of the most interesting integrals appearing in our calculations do not satisfy 
this condition. It can be applied, if e.g. quadratic forms of Feynman parameters can be transformed away or mapped into the argument 
of the iterated integral. In the massive case, however, this is not always possible, whatever order of integrations is applied.

Many of the most complicated integrals we have encountered so far were solved using the differential equations method. The idea behind this method is to take derivatives of the master integrals 
with respect to the invariants of the problem, and then re-express the result in terms of the master integrals themselves. This leads to a system of differential equations that can then be solved 
once appropriate boundary conditions are found. In our case, we take advantage of the introduction of the auxiliary variable $x$, as shown in 
Eq.~(\ref{trick}), and take derivatives with respect to this
variable. For example, consider the following two master integrals, which were needed to obtain $A_{Qq}^{(3), \rm PS}$,
\begin{eqnarray}
M_1(x) &=& \int \frac{d^D k_1}{(2\pi)^D} \frac{d^D k_2}{(2\pi)^D} \frac{d^D k_3}{(2\pi)^D}
\frac{1}{D_1 D_2 D_3 D_4 D_5 D_6 D_7}\, , \\
M_2(x) &=& \int \frac{d^D k_1}{(2\pi)^D} \frac{d^D k_2}{(2\pi)^D} \frac{d^D k_3}{(2\pi)^D}
\frac{1}{D_1^2 D_2 D_3 D_4 D_5 D_6 D_7} \, ,
\end{eqnarray}
where
\begin{eqnarray}
& D_1 = (k_1-p)^2, \quad D_2 = (k_2-p)^2, \quad D_3 = k_3^2-m^2, \quad D_4 = (k_1-k_3)^2-m^2, & \\
& \quad D_5 = (k_2-k_3)^2-m^2, \quad D_6 = 1 - x \Delta.k_3, \quad D_7 = 1 - x (\Delta.k_3-\Delta.k_1). &
\end{eqnarray}
Here $m$ is the mass of the heavy quark, and $p$ the momentum of the external light quark, which is taken on-shell ($p^2=0$). 
Taking derivatives with respect to $x$ we obtain
\begin{eqnarray}
\label{diffeq1}
\frac{d}{dx} M_1(x) &=& \frac{1}{1-x} \left(2+\epsilon -\frac{1}{x} \right) M_1(x)+\frac{2 x}{1-x} M_2(x) + \frac{K_1(x)}{1-x}, 
\\
\frac{d}{dx} M_2(x) &=& -\frac{1}{1-x} \left(\frac{1-2\epsilon}{x}+\frac{3}{2}\epsilon-2\right) M_2(x) \nonumber\\ &&
                         +\frac{\epsilon}{4}(2+3\epsilon) \frac{1}{1-x} \left( \frac{1}{x^2} - \frac{1}{x} \right) M_1(x) + 
\frac{K_2(x)}{1-x},
\label{diffeq2}
\end{eqnarray}
where $K_1(x)$ and $K_2(x)$ are linear combinations of sub-sector master integrals that have been solved previously. In 
Eqs.~(\ref{diffeq1},\ref{diffeq2}), we 
have set the mass $m$ and
$\Delta.p$ to 1 for simplicity. Now we undo the introduction of the variable $x$. Since
\begin{equation}
M_1(x) \propto \sum_{N=0}^{\infty} x^N F_1(N) \quad {\rm and} \quad M_2(x) \propto \sum_{N=0}^{\infty} x^N F_2(N),
\end{equation}
we obtain the following system of difference equations,
\begin{eqnarray}
(N+2) F_1(N+1) - (N+2+\epsilon) F_1(N) -2 F_2(N-1) &=& K_1(N) \, , \\
(N+2-2\epsilon) F_2(N+1) -\left( N+2-\frac{3}{2}\epsilon \right) F_1(N) && \\ 
   -\frac{\epsilon}{4}(2+3\epsilon) \left( F_1(N+2)-F_1(N+1) \right) &=& K_2(N) \, ,
\end{eqnarray}
where $K_1(N)$ and $K_2(N)$ are the $N$th terms of the Taylor expansions of $K_1(x)$ and $K_2(x)$, respectively. 
This system can now be solved using {\tt Sigma}, together with
the {\tt Mathematica} package {\tt OreSys} \cite{Gerhold:02};
for further details on this approach we refer to \cite{Bluemlein:2014qka}. In order to be able to do so, we need to obtain a 
few initial values for the integrals under consideration, which we can
do using the program {\tt MATAD} \cite{Steinhauser:2000ry} or by doing reductions of tensor integrals to scalar integrals \cite{Ablinger:2014uka}. Many of the master integrals needed to obtain 
$A_{Qq}^{(3), \rm PS}$ and some of the terms $\propto T_F^2$ in $A_{gg}^{(3)}$ were calculated this way. Recently, 3-loop quarkonic ladder 
and $V$-topology diagrams have also been obtained using this method, cf.~\cite{Bluemlein:2014qka}.
\section{Results}
\label{RES}

\vspace*{1mm}
\noindent
The expressions obtained for the operator matrix elements $A_{gq}^{(3)}$, $A_{qq}^{(3), \rm NS}$ and $A_{qq}^{(3), \rm TR}$ have been found to be 
given in terms of harmonic sums \cite{Vermaseren:1998uu,Blumlein:1998if} of up to weight five. For $A_{Qq}^{(3), \rm PS}$ for the first time 
generalized sums \cite{Moch:2001zr,Ablinger:2013cf} appear in the final answer,  namely,
\begin{equation}
S_{a,\vec{b}}(\zeta,\vec{\xi};N) 
= \sum_{k=1}^N \frac{\zeta^k}{k^a} S_{\vec{b}}(\vec{\xi};k)
\equiv S_{a,\vec{b}}(\zeta,\vec{\xi})~. 
\end{equation}
In particular, the constant term in $A_{Qq}^{(3), \rm PS}$ contains the following generalized sums
\begin{eqnarray*}
&& S_1\left(\frac{1}{2}\right), \,\, 
S_2\left(\frac{1}{2}\right), \,\,
S_3\left(\frac{1}{2}\right), \,\,
S_{1,1}\left(\frac{1}{2},1\right), \,\,
S_{1,1}\left(1,\frac{1}{2}\right), \,\,
S_{2,1}\left(\frac{1}{2},1\right), \,\,
S_{1,2}\left(\frac{1}{2},1\right), \,\,
S_{2,1}\left(1,\frac{1}{2}\right), \,\,  \\ &&
S_{1,2}\left(1,\frac{1}{2}\right), \,\,
S_{1,1,1}\left(\frac{1}{2},1,1\right), \,\,
S_{1,1,1}\left(1,\frac{1}{2},1\right), \,\,
S_{1,1,1}\left(1,1,\frac{1}{2}\right), \,\,
S_3\left(2\right), \,\,
S_{1,2}\left(2,1\right), \,\,
S_{2,1}\left(2,1\right),   
\end{eqnarray*}
etc., where we have omitted the explicit dependence on $N$. 
\begin{figure}[H]
\includegraphics[width=1\textwidth]{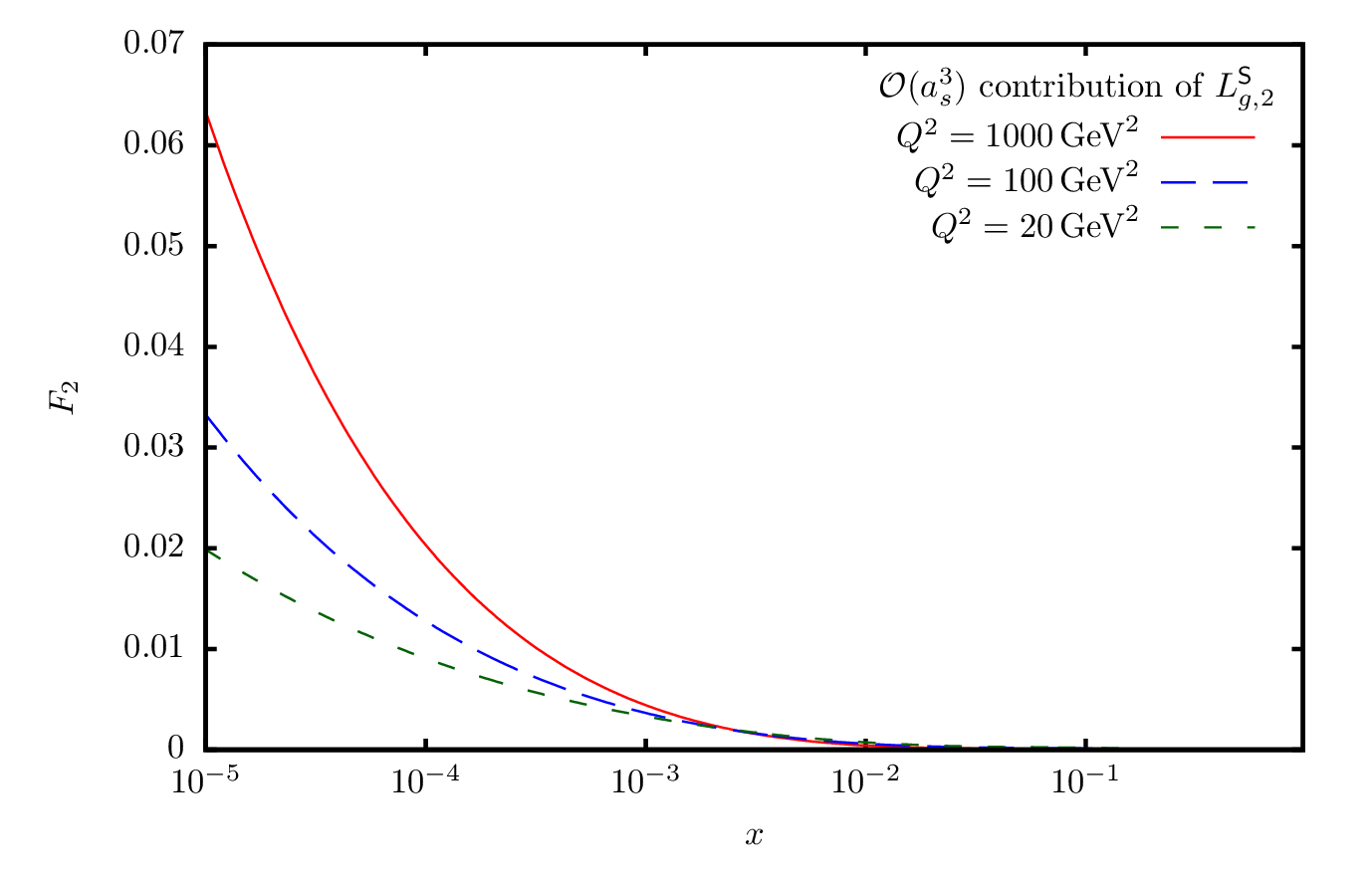}
\caption[]{\label{fig2} 
The $O(a_s^3)$ contribution by $L_{g,2}^{\sf S}$ to the structure function $F_2(x,Q^2)$ 
for $m_c = 1.59~{\rm GeV}$ using the parton distributions \cite{Alekhin:2013nda} 
(from Ref.~\cite{Behring:2014eya}).}
\end{figure}
In terms of a Mellin transform, 
\begin{equation}
\hat{f}(N) = \int_0^1 dx \,\, x^{N-1} f(x)
\end{equation}
these sums lead to generalized harmonic polylogarithms \cite{Ablinger:2013cf}. Using a recent reduction mechanism available in 
{\tt HarmonicSums} 
we were able to transform the physical result into the harmonic polylogarithms \cite{Remiddi:1999ew} evaluated at $x$ and $1 - 2 x$.
There are also other equivalent representations requiring generalizations of the Mellin transform, cf.~\cite{PS}.

In the case of the terms $\propto T_F^2$ in $A_{gg}^{(3)}$ \cite{Ablinger:2014uka} and for $V$-graph topologies contributing to 
$A_{Qg}^{(3)}$ \cite{Ablinger:2014yaa} we also found finite nested (inverse) binomial sums over (generalized) harmonic sums such as
\begin{equation}
\frac{1}{4^N} \binom{2 N}{N} \sum_{k=0}^N \frac{4^k}{k^l\binom{2 k}{k}} S_1(k),~~~l \in \mathbb{N}
\end{equation}
or
\begin{equation}
\sum_{i=1}^N \binom{2 i}{i} (-2)^i \sum_{j=1}^i \frac{1}{j \binom{2 j}{j}} S_{1,2}\left(\frac{1}{2},-1;j\right),
\end{equation}
where $S_{\vec{a}}(N)$ denotes a nested harmonic sum.

Doing the inverse Mellin transform of these sums we find that these are expressed in terms of iterated integrals over 
root-valued
alphabets. In total, we have found that 33 new letters are needed in the algebraically irreducible representations for the calculations we 
have done so far.

The calculation of all these OMEs has allowed us also to check the corresponding contributions to the 3-loop anomalous dimensions. We
find perfect agreement with the literature. In the case of transversity, these have been calculated for the first time ab initio. 
\begin{figure}[H]
\includegraphics[width=1\textwidth]{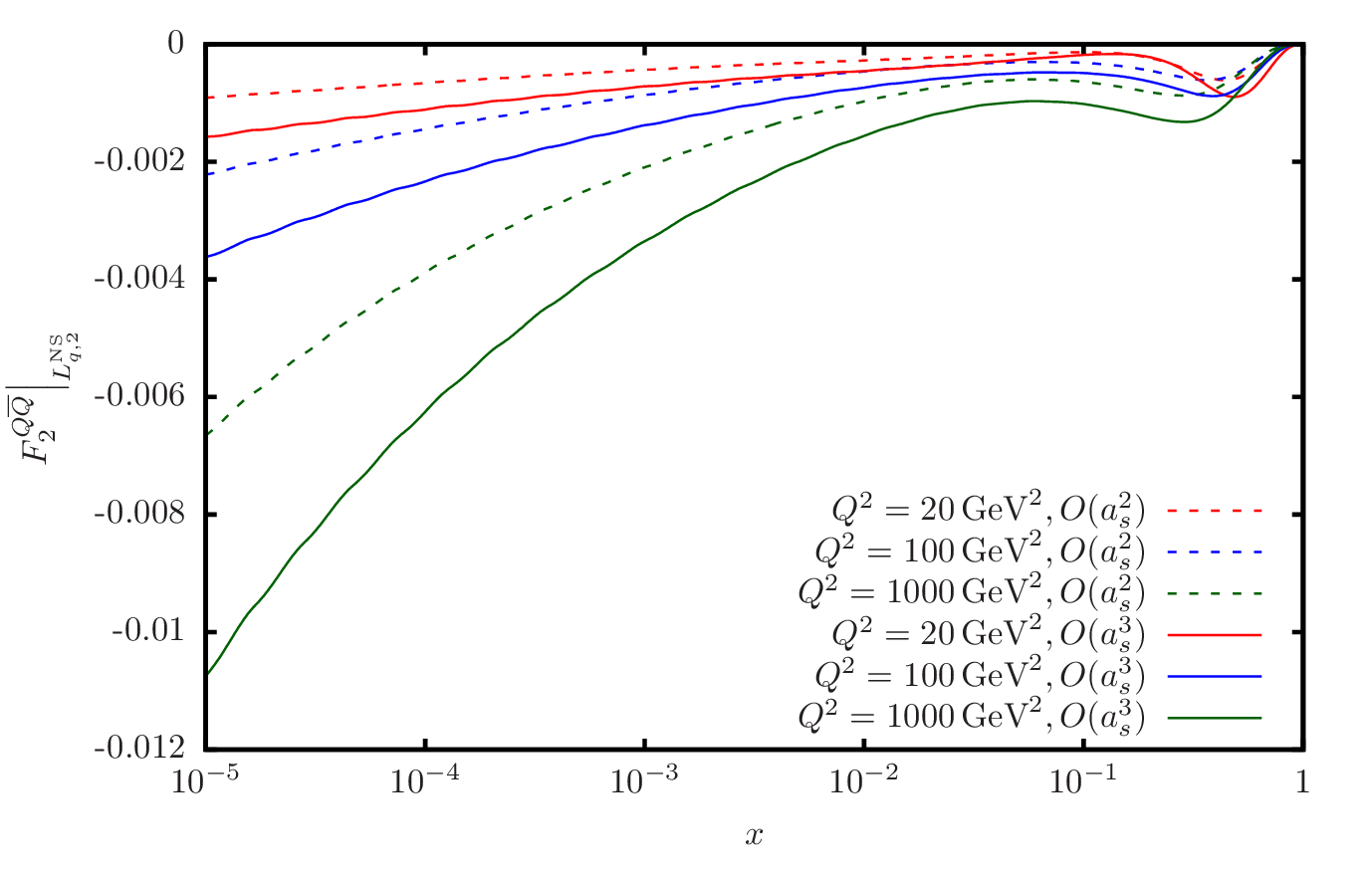}
\caption{\small The flavor non-singlet contribution of the Wilson coefficient $L^{\rm NS}_{q,2}$ to the structure function
$F_2(x,Q^2)$ at 2- and 3-loop order using the NNLO parton distribution functions \cite{Alekhin:2013nda} in the on-shell scheme for $m_c=1.59$ 
GeV (from Ref.~\cite{Ablinger:2014vwa}).}
\label{NSgraph}
\end{figure}

Having calculated these OMEs, the remaining tasks are the convolution with the massless Wilson coefficients and then with the PDFs, in order 
to obtain the contributions to the structure functions. We have obtained numerical results for the Wilson coefficients $L_{q,(2)}^{\rm 
PS}$, $L_{g,(2)}^{\rm S}$ \cite{Behring:2014eya} and $L_{q,(2)}^{\rm NS}$ \cite{Ablinger:2014vwa} to 3-loop order. 
In Figure~\ref{fig2} the 3-loop corrections by the Wilson coefficient $L_{g,2}^{S}$ is shown. In the kinematic region probed 
by HERA it reaches $\sim 1\%$, i.e. the experimental accuracy and is therefore of importance. They are larger than the 2-loop 
corrections for this quantity, due to a term $\propto 1/z$ emerging first in the 3-loop corrections, cf. \cite{Behring:2014eya}.
In Figure~\ref{NSgraph}, we show the contribution of the heavy flavor non-singlet Wilson coefficient to structure function $F_2(x,Q^2)$ at 2- 
and 3-loop order, for different values of $Q^2$. They turn out to be smaller than $1\%$ in the kinematic region of HERA. In 
Ref.~\cite{Ablinger:2014vwa} we also presented the complete transformation coefficients in the VFNS in the non-singlet case at 
3-loop order.
Future high-luminosity machines such as the EIC \cite{Boer:2011fh} will reach a much higher resolution for $F_2(x,Q^2)$ than HERA. Here all 
these terms will be of experimental relevance. Numerical results on the pure-singlet contributions will be given later this year.
\section{Conclusions}
\label{CONC}

\vspace*{1mm}
\noindent
Considerable progress has been made recently in the calculation of the NNLO heavy flavor contributions to the structure functions in DIS 
for large values of $Q^2$. By now, six out of eight operator matrix elements (and the associated anomalous dimensions) have been completed, 
and partial results are available for the remaining two OMEs. This progress was possible thanks to the development of new computer algebra 
and mathematical technologies. Several programs have played a crucial role in these calculations, such as {\tt Reduze2} for the reduction to 
master integrals, and {\tt Sigma}, {\tt HarmonicSums}, {\tt EvaluateMultiSums}, {\tt SumProduction} and {\tt OreSys} for summation algorithms 
and the solution of difference equations. These programs and the algorithms associated with them continue to be developed and refined as we 
encounter ever more challenging problems in this endeavor. The 3-loop heavy flavor Wilson coefficients calculated so far yield 
contributions to $F_2(x,Q^2)$ of $O(\lesssim 1\%)$, cf.~\cite{Behring:2014eya,Ablinger:2014vwa}, reaching the experimental accuracy of the 
structure function $F_2(x,Q^2)$ at HERA. We will report on numerical results for further Wilson coefficients and OMEs in the future. 
The completion of this project is underway and will allow us to make more precise determinations of $\alpha_s$ and $m_c$, the parton 
distribution functions, as well as to establish the VFNS at NNLO, needed for predictions at hadron colliders such as the LHC.



\begin{thebibliography}{99}
%
\bibitem{BMSN96}
  M.~Buza, Y.~Matiounine, J.~Smith, R.~Migneron and W.~L.~van Neerven,
  Nucl.\ Phys.\ B {\bf 472} (1996) 611
  [hep-ph/9601302].
%
\bibitem{BBK09NPB}
  I.~Bierenbaum, J.~Bl\"umlein and S.~Klein,
  {Nucl.\ Phys.}\ B {\bf 820} (2009) 417
  [arXiv:0904.3563 [hep-ph]].
%
\bibitem{MVV2005}
  J.A.M.~Vermaseren, A.~Vogt and S.~Moch,
  Nucl.\ Phys.\ B {\bf 724} (2005) 3
  [hep-ph/0504242].
%
\bibitem{ABKSW11NPB}
  J.~Ablinger, J.~Bl\"umlein, S.~Klein, C.~Schneider and F.~Wi\ss{}brock,
  Nucl.\ Phys.\ B {\bf 844} (2011) 26
  [arXiv:1008.3347 [hep-ph]].
%
\bibitem{Ablinger:2014lka}
  J.~Ablinger, J.~Bl\"umlein, A.~De Freitas, A.~Hasselhuhn, A.~von Manteuffel, M.~Round, C.~Schneider and F.~Wi\ss{}brock,
  Nucl.\ Phys.\ B {\bf 882} (2014) 263
  [arXiv:1402.0359 [hep-ph]].
%
\bibitem{Ablinger:2014vwa}
  J.~Ablinger, A.~Behring, J.~Bl\"umlein, A.~De Freitas, A.~Hasselhuhn, A.~von Manteuffel, M.~Round, C.~Schneider, and F.~Wi\ss{}brock,
  arXiv:1406.4654 [hep-ph], Nucl. Phys. {\bf B} (2014) in press.
%
\bibitem{PS}
J.~Ablinger et al.,  DESY 13--232.
%
\bibitem{Behring:2014eya}
  A.~Behring, I.~Bierenbaum, J.~Bl\"umlein, A.~De~Freitas, S.~Klein and F.~Wi\ss{}brock,
  arXiv:1403.6356 [hep-ph].
%
\bibitem{Blumlein:2006mh}
  J.~Bl\"umlein, A.~De Freitas, W.~L.~van Neerven and S.~Klein,
  Nucl.\ Phys.\ B {\bf 755} (2006) 272
  [hep-ph/0608024].
%
\bibitem{Ablinger:2014uka}
  J.~Ablinger, J.~Bl\"umlein, A.~De Freitas, A.~Hasselhuhn, A.~von Manteuffel, M.~Round and C.~Schneider,
  Nucl. Phys. B {\bf 885} (2014) 280 [arXiv:1405.4259 [hep-ph]].
%
\bibitem{Blumlein:2012vq}
  J.~Bl\"umlein, A.~Hasselhuhn, S.~Klein and C.~Schneider,
  Nucl.\ Phys.\ B {\bf 866} (2013) 196
  [arXiv:1205.4184 [hep-ph]].
%
\bibitem{Behring:2013dga}
  A.~Behring, J.~Bl\"umlein, A.~De Freitas, T.~Pfoh, C.~Raab, M.~Round, J.~Ablinger, A.~Hasselhuhn, C.~Schneider, and F.~Wi\ss{}brock.
  PoS RADCOR {\bf 2013} (2014) 058
  [arXiv:1312.0124 [hep-ph]].
%
\bibitem{Alekhin:2012vu}
  S.~Alekhin, J.~Bl\"umlein, K.~Daum, K.~Lipka and S.~Moch,
  Phys.\ Lett.\ B {\bf 720} (2013) 172
  [arXiv:1212.2355 [hep-ph]].
%
\bibitem{PDF}
  P.~Jimenez-Delgado and E.~Reya,
  Phys.\ Rev.\ D {\bf 89} (2014) 074049
  [arXiv:1403.1852 [hep-ph]];
  R.~S.~Thorne,
  arXiv:1402.3536 [hep-ph];\\
  R.~D.~Ball {\it et al.}  [NNPDF Collaboration],
  Nucl.\ Phys.\ B {\bf 877} (2013) 290
  [arXiv:1308.0598 [hep-ph]];\\
  J.~Gao, M.~Guzzi, J.~Huston, H.~-L.~Lai, Z.~Li, P.~Nadolsky, J.~Pumplin and D.~Stump {\it et al.},
  Phys.\ Rev.\ D {\bf 89} (2014) 033009
  [arXiv:1302.6246 [hep-ph]].
%
\bibitem{Alekhin:2013nda}
  S.~Alekhin, J.~Bl\"umlein and S.~Moch,
  Phys.\ Rev.\ D {\bf 89} (2014) 054028
  [arXiv:1310.3059 [hep-ph]].
%
\bibitem{Buza:1996wv}
  M.~Buza, Y.~Matiounine, J.~Smith and W.~L.~van Neerven,
  Eur.\ Phys.\ J.\ C {\bf 1} (1998) 301
  [hep-ph/9612398].
%
\bibitem{JB14a}
J. Bl\"umlein and F. Wi\ss{}brock, DESY 14--019.
%
\bibitem{Ablinger:2011pb}
  J.~Ablinger, J.~Bl\"umlein, S.~Klein, C.~Schneider and F.~Wi\ss{}brock,
  arXiv:1106.5937 [hep-ph].
%
\bibitem{Ablinger:2012qj}
  J.~Ablinger, J.~Bl\"umlein, A.~Hasselhuhn, S.~Klein, C.~Schneider and F.~Wi\ss{}brock,
  PoS RADCOR2011 (2011) 031
  [arXiv:1202.2700 [hep-ph]].
%
\bibitem{Ablinger:2014fla}
  J.~Ablinger, J.~Bl\"umlein, A.~De Freitas, A.~Hasselhuhn, A.~von Manteuffel, M.~Round, C.~Schneider and F.~Wi\ss{}brock,
  arXiv:1407.2821 [hep-ph].
%
\bibitem{Buza:1997mg}
  M.~Buza and W.~L.~van Neerven,
  Nucl.\ Phys.\ B {\bf 500} (1997) 301
  [hep-ph/9702242].
%
\bibitem{Blumlein:2014fqa}
  J.~Bl\"umlein, A.~Hasselhuhn and T.~Pfoh,
  Nucl.\ Phys.\ B {\bf 881} (2014) 1
  [arXiv:1401.4352 [hep-ph]].
%
\bibitem{Klein:2009ig}
  S.W.G.~Klein,
  arXiv:0910.3101 [hep-ph].
%
\bibitem{Nogueira:1991ex}
  P.~Nogueira, 
  J.\ Comput.\ Phys.\  {\bf 105} (1993) 279.
%
\bibitem{FORM}
  J.A.M.~Vermaseren,
  math-ph/0010025;\\
  M.~Tentyukov and J.A.M.~Vermaseren,
  Comput.\ Phys.\ Commun.\  {\bf 181} (2010) 1419
  [hep-ph/0702279].
%
\bibitem{vonManteuffel:2012np}
  A.~von Manteuffel and C.~Studerus,
  arXiv:1201.4330 [hep-ph];\\
  C.~Studerus,
  Comput.\ Phys.\ Commun.\  {\bf 181} (2010) 1293
  [arXiv:0912.2546 [physics.comp-ph]].
%
\bibitem{FERMAT}
R.H.~Lewis, Computer Algebra System {\tt Fermat}, {\tt http://home.bway.net/lewis}.
%
\bibitem{Bauer:2000cp}
  C.~W.~Bauer, A.~Frink and R.~Kreckel,
  Symbolic Computation {\bf 33} (2002) 1,
  cs/0004015 [cs-sc].
%
\bibitem{Laporta:2001dd}
  S.~Laporta,
  Int.\ J.\ Mod.\ Phys.\ A {\bf 15} (2000) 5087
  [hep-ph/0102033].
%
\bibitem{SIG1}
C.~Schneider, {S\'em.~Lothar. Combin.\/} {\bf 56} (2007) 1, 
 article B56b.
%
\bibitem{SIG2}
C.~Schneider, {{Computer Algebra in Quantum Field Theory: Integration,
  Summation and Special Functions}\/} Texts and Monographs in Symbolic
  Computation eds. C.~Schneider and J.~Bl\"umlein  (Springer, Wien, 2013) 325 
  arXiv:1304.4134 [cs.SC].
%
\bibitem{Karr:81}
M.~Karr 1981 {J.~ACM\/} {\bf 28} (1981) 305.
%
\bibitem{Schneider:01}
C.~Schneider, 
{\it Symbolic Summation in Difference Fields\/} Ph.D. Thesis
RISC, Johannes Kepler University, Linz technical report 01-17 (2001).
%
\bibitem{Schneider:05a}
C.~Schneider, {J. Differ. Equations Appl.\/} {\bf 11} (2005) 799.
%
\bibitem{Schneider:07d}
C.~Schneider, {J. Algebra Appl.\/} {\bf 6} (2007) 415. 
%
\bibitem{Schneider:08c}
C.~Schneider, {J. Symbolic Comput.\/} {\bf 43} (2008) 611 
  [arXiv:0808.2543].
%
\bibitem{Schneider:10a}
C.~Schneider, {Appl. Algebra Engrg. Comm. Comput.\/} {\bf 21} (2010) 1. 
%
\bibitem{Schneider:10b}
C.~Schneider, {\it {Motives, Quantum Field Theory, and Pseudodifferential
  Operators}\/} ({Clay Mathematics Proceedings\/} Vol.~{\bf 12}, eds. A.~Carey,
  D.~Ellwood, S.~Paycha and S.~Rosenberg (Amer. Math. Soc) (2010), 285 
  arXiv:0808.2543.
%
\bibitem{Schneider:10c}
C.~Schneider, {Ann. Comb.\/} {\bf 14} (2010)  533 
[arXiv:0808.2596].
%
\bibitem{Schneider:13b}
C.~Schneider,
in~: Lecture Notes in Computer Science (LNCS)
eds. J. Guitierrez, J. Schicho, M.~Weimann, in press,
arXiv:1307.7887 [cs.SC] (2013).
%
\bibitem{Ablinger:2010kw}
  J.~Ablinger,
  arXiv:1011.1176 [math-ph].
%
\bibitem{Ablinger:2013hcp}
  J.~Ablinger,
  arXiv:1305.0687 [math-ph].
%
\bibitem{Ablinger:2011te}
  J.~Ablinger, J.~Bl\"umlein and C.~Schneider,
  J.\ Math.\ Phys.\  {\bf 52} (2011) 102301
  [arXiv:1105.6063 [math-ph]] and in preparation.
%
\bibitem{Ablinger:2013cf}
  J.~Ablinger, J.~Bl\"umlein and C.~Schneider,
  J.\ Math.\ Phys.\  {\bf 54} (2013) 082301
  [arXiv:1302.0378 [math-ph]].
%
\bibitem{EMSSP}
  J.~Ablinger, J.~Bl\"umlein, S.~Klein and C.~Schneider,
  Nucl.\ Phys.\ Proc.\ Suppl.\  {\bf 205-206} (2010) 110
  [arXiv:1006.4797 [math-ph]];\\
  J.~Bl\"umlein, A.~Hasselhuhn and C.~Schneider,
  PoS RADCOR {\bf 2011} (2011) 032
  [arXiv:1202.4303 [math-ph]];\\
\bibitem{Schneider:2013zna}
  C.~Schneider,
  J.\ Phys.\ Conf.\ Ser.\  {\bf 523} (2014) 012037
  [arXiv:1310.0160 [cs.SC]].
%
\bibitem{HYPERGEOM}
W.N. Bailey, {\it Generalized Hypergeometric Series}, (Cambridge University
Press,  Cambridge, 1935);\\
L.J. Slater, {\it Generalized Hypergeometric Functions}, (Cambridge University
Press, Cambridge, 1966);\\
P. Appell and J. Kamp\'{e} de F\'{e}riet, {\it Fonctions
Hyperg\'{e}om\'{e}triques et Hypersp\'{e}riques, Polynomes D' Hermite},
(Gauthier-Villars, Paris, 1926);\\
P. Appell, {\it Les Fonctions Hyperg\"{e}om\'{e}triques de Plusieur
Variables}, (Gauthier-Villars, Paris, 1925);\\
J. Kamp\'{e} de F\'{e}riet, {\sf La fonction
hyperg\"{e}om\'{e}trique},(Gauthier-Villars, Paris, 1937);\\
H. Exton, {\it Multiple Hypergeometric Functions and Applications},
(Ellis Horwood, Chichester, 1976);\\
H. Exton, {\it Handbook of Hypergeometric Integrals},
(Ellis Horwood, Chichester, 1978);\\
H.M. Srivastava and P.W. Karlsson, {\it Multiple Gaussian Hypergeometric
Series}, (Ellis Horwood, Chicester, 1985).
%
\bibitem{TWOL}
  I.~Bierenbaum, J.~Bl\"umlein and S.~Klein,
  Nucl.\ Phys.\ B {\bf 780} (2007) 40
  [hep-ph/0703285];\\
  I.~Bierenbaum, J.~Bl\"umlein, S.~Klein and C.~Schneider,
  Nucl.\ Phys.\ B {\bf 803} (2008) 1
  [arXiv:0803.0273 [hep-ph]];\\
  I.~Bierenbaum, J.~Bl\"umlein and S.~Klein,
  Phys.\ Lett.\ B {\bf 672} (2009) 401
  [arXiv:0901.0669 [hep-ph]].
%
\bibitem{Ablinger:2012qm}
  J.~Ablinger, J.~Bl\"umlein, A.~Hasselhuhn, S.~Klein, C.~Schneider and F.~Wi\ss{}brock,
  Nucl.\ Phys.\ B {\bf 864} (2012) 52
  [arXiv:1206.2252 [hep-ph]].
%
\bibitem{Mellin1895}
H.~Mellin,
{Acta Societatis Scientiarum Fennicae}, {\bf XX.}(7) (1895)
  1; 
Math. Ann. {\bf 68} (1910) 305.
%
\bibitem{Barnes1908}
E.W. Barnes, Proc. Lond. Math. Soc. (2) {\bf 6} (1908) 141; Quart. J.
Math. {\bf 41} (1910) 136.
%
\bibitem{Smirnov:1999gc}
  V.~A.~Smirnov,
  Phys.\ Lett.\ B {\bf 460} (1999) 397
  [hep-ph/9905323].
%
\bibitem{Smirnov2006}
V.~A. Smirnov, {\it {Feynman Integral Calculus}}, (Springer, Berlin, 2006).
%
\bibitem{Brown:2008um}
  F.~Brown,
  Commun.\ Math.\ Phys.\  {\bf 287} (2009) 925
  [arXiv:0804.1660 [math.AG]].
%
\bibitem{Ablinger:2014yaa}
  J.~Ablinger, J.~Bl\"umlein, C.~Raab, C.~Schneider and F.~Wi\ss{}brock,
  Nucl. Phys. {\bf B885} (2014) 409  arXiv:1403.1137 [hep-ph].
%
\bibitem{DEQ}
  A.~V.~Kotikov,
  Phys.\ Lett.\ B {\bf 254} (1991) 158;\\
  M.~Caffo, H.~Czyz, S.~Laporta and E.~Remiddi,
  Acta Phys.\ Polon.\ B {\bf 29} (1998) 2627
  [hep-th/9807119];\\
  Nuovo Cim.\ A {\bf 111} (1998) 365
  [hep-th/9805118];\\
  T.~Gehrmann and E.~Remiddi,
  Nucl.\ Phys.\ B {\bf 580} (2000) 485
  [hep-ph/9912329];\\
  M.~Caffo, H.~Czyz and E.~Remiddi,
  Nucl.\ Phys.\ B {\bf 634} (2002) 309
  [hep-ph/0203256].
%
\bibitem{Bluemlein:2014qka}
  J.~Bl\"umlein, A.~De Freitas and C.~Schneider,
  arXiv:1407.2537 [cs.SC], PoS (LL2014) 017.
%
\bibitem{ABLINGER:14}
J.~Ablinger, 
PoS (LL2014) 019.
%
\bibitem{RAAB:14}
J.~Ablinger, J.~Bl\"umlein, C.~Raab, and C. Schneider, 
PoS (LL2014) 020.
%
\bibitem{Ablinger:2014bra}
  J.~Ablinger, J.~Bl\"umlein, C.~G.~Raab and C.~Schneider,
  arXiv:1407.1822 [hep-th].
%
\bibitem{Ablinger:2013jta}
  J.~Ablinger and J.~Bl\"umlein,
  arXiv:1304.7071 [math-ph].
%
\bibitem{Czakon:2005rk}
  M.~Czakon,
  Comput.\ Phys.\ Commun.\  {\bf 175} (2006) 559
  [hep-ph/0511200].
%
\bibitem{Gerhold:02}
S.~Gerhold,
{\it Uncoupling systems of linear ore operator equations},
\newblock Master's thesis, RISC, J.~Kepler University, Linz, 2002.
%
\bibitem{Steinhauser:2000ry}
  M.~Steinhauser,
  Comput.\ Phys.\ Commun.\  {\bf 134} (2001) 335
  [hep-ph/0009029].
%
\bibitem{Vermaseren:1998uu}
  J.A.M.~Vermaseren,
  Int.\ J.\ Mod.\ Phys.\  A {\bf 14} (1999) 2037
  [arXiv:hep-ph/9806280].
%
\bibitem{Blumlein:1998if}
  J.~Bl\"umlein and S.~Kurth,
  Phys.\ Rev.\  D {\bf 60} (1999) 014018
  [arXiv:hep-ph/9810241].
%
\bibitem{Moch:2001zr}
  S.~Moch, P.~Uwer and S.~Weinzierl,
  J.\ Math.\ Phys.\  {\bf 43} (2002) 3363
  [hep-ph/0110083].
%
\bibitem{Remiddi:1999ew}
  E.~Remiddi and J.~A.~M.~Vermaseren,
  Int.\ J.\ Mod.\ Phys.\ A {\bf 15} (2000) 725
  [hep-ph/9905237].
%
\bibitem{Boer:2011fh}
  D.~Boer, M.~Diehl, R.~Milner, R.~Venugopalan, W.~Vogelsang, D.~Kaplan, H.~Montgomery and S.~Vigdor {\it et al.},
  arXiv:1108.1713 [nucl-th];\\
  A.~Accardi, J.~L.~Albacete, M.~Anselmino, N.~Armesto, E.~C.~Aschenauer, A.~Bacchetta, D.~Boer and W.~Brooks {\it et al.},
  arXiv:1212.1701 [nucl-ex].
\end{thebibliography}
\end{document}